# Initial Evaluation of the Usability of Front-End Ontology Tooling

Clair Kronk[1,2,*,†] and Rishabh Jain[1,†]

[1] Institute for Health Equity Research (IHER), Department of Population Health Science & Policy (PHSP), Icahn School of Medicine at Mount Sinai (ISMMS), New York, NY, USA

[2] Department of Artificial Intelligence and Human Health (AIHH), Icahn School of Medicine at Mount Sinai (ISMMS), New York, NY, USA

**Abstract**

Introduction. Ontologies are widely used biomedical science and clinical practice. However, no recent works have analyzed the usability of ontology development software.
Methods. We survey ontology researchers to assess the usability of 15 front-end ontology tools using the System Usability Scale (SUS).
Results. Among 38 respondents, Protégé and WebProtégé were most used but showed only moderate usability (SUS ≈ 60). Familiarity significantly predicted usability scores (p=0.016).
Discussion. Results highlight a usability gap in ontology tooling critical for advancing biomedical data integration.

## 1. Introduction

Ontologies have become essential in biomedical science and clinical practice, enabling the findability, accessibility, interoperability, and reusability of data across sites and disciplines [2, 8]. Ontologies are formal, structured representations of concepts and relationships within a domain, serving as critical tools for ensuring data consistency and supporting automated reasoning [6, 14, 15]. For example, ontologies are used to harmonize electronic health records (EHRs), create diagnostic algorithms, and standardize public health recommendations [3, 9, 13, 16]. When ontologies are flawed, the consequences can range from increased risks of medication errors, fragmentation of patient data, harm from redundant testing or misdiagnosis, insurance claim denials, and additional healthcare expenditure [1, 4, 5, 12, 18, 19]. Such instability can stem from inadequate reasoning validation and provenance control in current ontology tool environments, which could be impacted by the usability of such tooling [7, 10]. To assess the usability of current ontology development software, we conducted an initial survey of ontology researchers regarding front-end ontology tooling software.

## 2. Methods

In October 2025, we sent an IRB-approved RedCAP survey to three online groups of ontology researchers. This survey asked researchers their level of familiarity with ontologies generally, as well as familiarity with each of 15 different identified front-end ontology development software solutions, being Apollo, Fluent Editor, Hozo, NeOn Toolkit, OboEdit, OntoGraf, OntoStudio, OwlGred, PoolParty, Protégé, Swoop, TopBraid Composer, Vitro, VocBench, and WebProtégé. Users were also asked if they utilized another tool not listed amongst these. After each tool, users were asked qualitatively about the strengths and weaknesses of each tool, and at the end of the survey, about the various strengths and limitations of current software solutions as a whole. Descriptive statistics were calculated for all 15 ontology development tools, including usage frequency (n, %), mean and standard deviations for familiarity ratings (4-point scale), each of the 10 individual System Usability Scale (SUS) items, and overall SUS scores. SUS scores were calculated using standard methodology [11], yielding values from 0-100 for each tool-respondent combination. SUS scores were compared across tools using ANCOVA analysis, adjusting for familiarity as a covariate. Post-hoc pairwise comparisons between tools were performed using Tukey's Honestly Significant Difference test using robust standard errors clustered by respondent to account for within-subject correlation. Additionally, independent samples t-tests (Welch's t-tests in case of unequal variances) compared SUS scores between high familiarity (ratings 3-4) and low familiarity (ratings 1-2) groups, separately for each tool with ≥5 sample size per group. Effect sizes were reported using Cohen's d and statistical significance was set at $\alpha = 0.05$. Most common n-grams (1, 2, 3) were determined from the strengths and limitations after normalizing case and removing stopwords. Sentiment analysis was performed for both strengths and limitations using the Natural Language Tool Kit (NLTK) sentiment intensity analyzer trained using the VADER lexicon.

* Corresponding author.
† These authors contributed equally.
Clair.Kronk@mountsinai.org (C. Kronk); j.rishabh95@yahoo.com (R. Jain)
0000-0001-8397-8810 (C. Kronk)
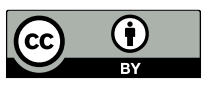

## 3. Results

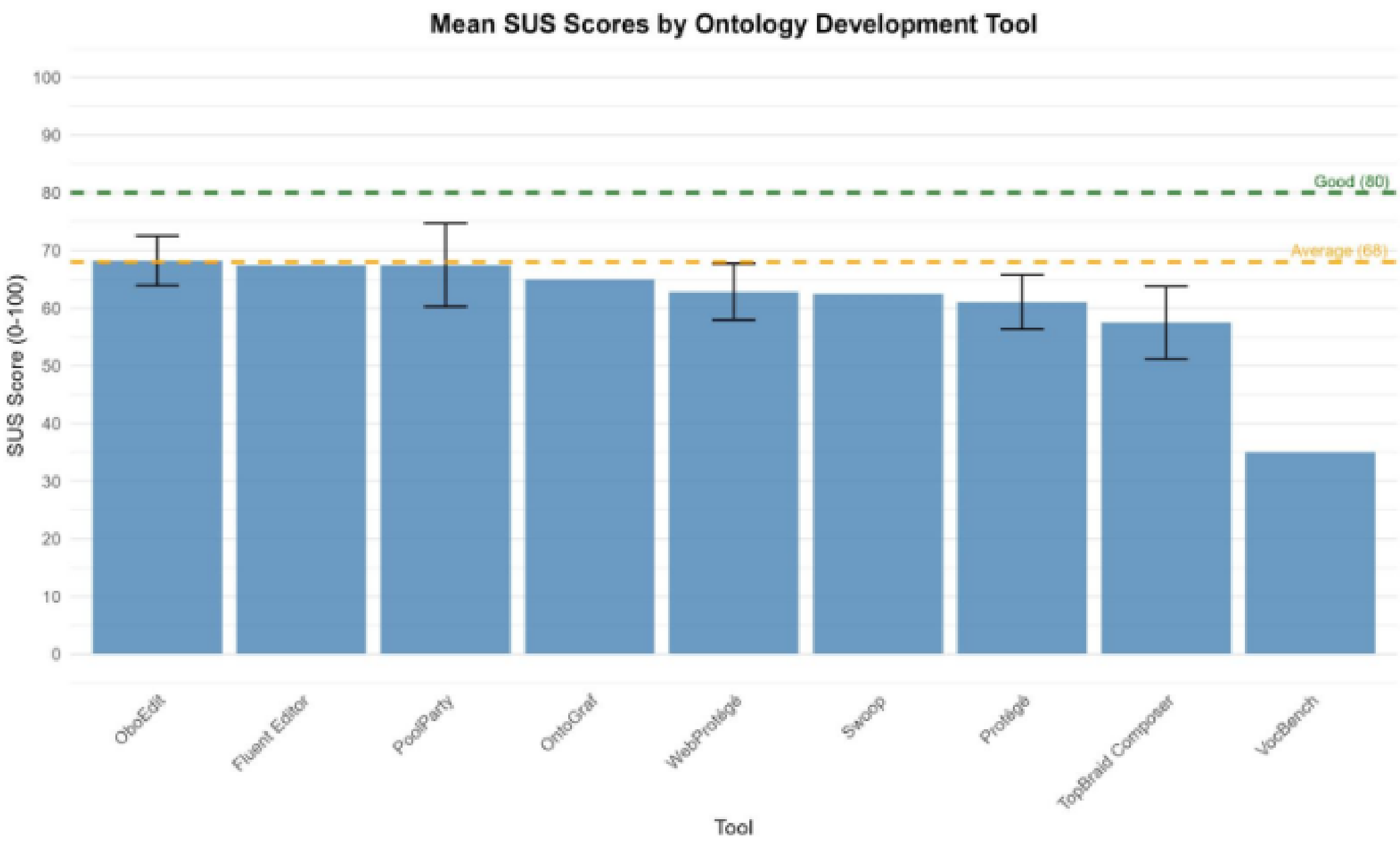

**Figure 1:** Mean Software Usability Scale (SUS) scores for individual tools organized from highest to lowest average score. Standard deviations shown where able to be calculated.

* Corresponding author.
† These authors contributed equally.
Clair.Kronk@mountsinai.org (C. Kronk); j.rishabh95@yahoo.com (R. Jain)
0000-0001-8397-8810 (C. Kronk)

Of 38 survey respondents, 28 (74%) used at least one of 15 ontology development tools. Protégé was the most widely used tool (92.9%, n=26, Mean SUS=61.06±23.83), followed by WebProtégé (53.6%, n=15, Mean SUS=62.83±18.85) and OboEdit (25.0%, n=7, Mean SUS=68.21±11.34), with most tools receiving Grade C (Mean SUS 65-71) or D (Mean SUS 51.7-62.6) ratings [17]. ANCOVA showed no significant differences in SUS scores between tools (p=0.659), but user familiarity significantly predicted SUS scores (p=0.016), with higher familiarity being associated with better SUS scores. Post-hoc pairwise comparisons showed no significant differences between any tool pairs. Tool-specific analysis was conducted only for Protégé, which had sufficient sample sizes in both groups. High familiarity users had higher SUS score (n=16, Mean SUS=66.56) compared to low familiarity users (n=6, Mean SUS=48.33), with a mean difference of 18.23 points (p=0.098, Cohen's d=0.83). Although not statistically significant, it represents a large effect size. The most common n-grams regarding across all strengths (not including those mentioning the names of tools themselves) included 'open-source' (n=8), 'powerful' (n=6), 'free' (n=5), 'integration' (n=5), 'simple' (n=3), and 'easy to use' (n=3). Limitations included terms like 'support' (n=10), 'load' (n=7), 'difficult' (n=6), and 'clunky' (n=5). Sentiment score across all strengths was 2.6% negative, 76.1% neutral, and 21.4% positive, whereas the sentiment score across all limitations was 8% negative, 77.9% neutral, and 14.1% positive.

**Table 1**
Descriptive statistics and SUS scores for Ontology Development Tools. *n* (%) represents absolute count of respondents using the tool and percentage across respondents using at least one tool (N=28). Familiarity and SUS scores are reported as Mean (SD) values. Standard deviations not valid when only 1 respondent uses the tool.

| Tool | *n* (%) | Familiarity | SUS Score | Grade |
|---|---|---|---|---|
| OboEdit | 7 (25.0%) | 2.67 (1.21) | 68.21 (11.34) | C |
| FluentEditor | 1 (3.6%) | 1.00 (NA) | 67.50 (NA) | C |
| PoolParty | 3 (10.7%) | 1.00 (0.00) | 67.50 (12.50) | C |
| OntoGraf | 1 (3.6%) | 1.00 (NA) | 65.00 (NA) | C |
| WebProtégé | 15 (53.6%) | 2.15 (0.90) | 62.83 (18.85) | C- |
| SWOOP | 1 (3.6%) | NA (NA) | 62.50 (NA) | D |
| Protégé | 26 (92.9%) | 3.14 (0.94) | 61.06 (23.83) | D |
| TopBraid Composer | 6 (21.4%) | 1.60 (0.89) | 57.50 (15.41) | D |
| VocBench | 1 (3.6%) | 1.00 (NA) | 35.00 (NA) | F |

## 4. Discussion

Ontology researchers generally indicated that software solutions are not very usable, with the possible exceptions of OboEdit and PoolParty, although our sample size was too small to say for certain. The strong effect of user familiarity on SUS scores indicates that perceptions of usability may be more influenced by technical expertise rather than interface design. The free and open-source nature of tools like Protégé was frequently cited as a strength, reflecting its continued dominance in ontology development. However, qualitative feedback highlighted issues with performance, intuitiveness of interfaces, and integration with modern workflows. Taken together, these findings indicate a gap between the functionality provided by current ontology tooling and the evolving needs of biomedical researchers.

* Corresponding author.
† These authors contributed equally.
Clair.Kronk@mountsinai.org (C. Kronk); j.rishabh95@yahoo.com (R. Jain)
0000-0001-8397-8810 (C. Kronk)
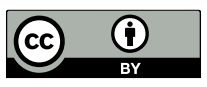

## Acknowledgements

Work by Clair Kronk and Rishabh Jain for this study was funded by NIH/NCI U54CA267776. This work was also supported in part through the resources and staff expertise provided by Scientific Computing and Data at the Icahn School of Medicine at Mount Sinai by the Clinical and Translational Science Awards (CTSA) grant ULTR004419 from the National Center for Advancing Translational Sciences.

* Corresponding author.
† These authors contributed equally.
Clair.Kronk@mountsinai.org (C. Kronk); j.rishabh95@yahoo.com (R. Jain)
0000-0001-8397-8810 (C. Kronk)

* Corresponding author.
† These authors contributed equally.
Clair.Kronk@mountsinai.org (C. Kronk); j.rishabh95@yahoo.com (R. Jain)
0000-0001-8397-8810 (C. Kronk)
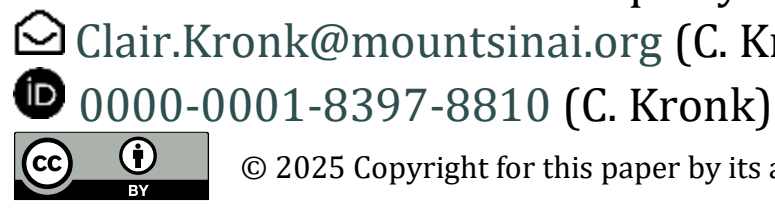